\documentclass[10pt]{article}
\usepackage{enumerate}
\usepackage{subcaption}
\usepackage[letterpaper]{geometry}
\usepackage{hicss}
\usepackage{times}
\usepackage[none]{hyphenat}
\usepackage{url}
\usepackage{latexsym}
\usepackage{indentfirst}
\usepackage{graphicx}
\graphicspath{{images/}}

\usepackage[
    style=apa,
    sorting=nyt,      
    uniquename=false,  
]{biblatex}
\let\cite\parencite
\usepackage{amsmath}
\usepackage{amsfonts}
\usepackage{amsthm}
\newtheorem{theorem}{Theorem}

\newtheorem{proposition}{Proposition}
\newtheorem{remark}{Remark}
\newtheorem{definition}{Definition}

\usepackage{algorithm}
\usepackage{algpseudocode}

\usepackage{newtxtext,newtxmath}
\usepackage{enumitem}
\setlist[enumerate]{leftmargin=*, itemsep=-.3em, topsep=3pt}

\allowdisplaybreaks[4]
\usepackage{xcolor, bm}
\newcommand{\blue}[1]{\textcolor{black}{#1}}

\usepackage{xurl}

\usepackage{hyperref}
\hypersetup{
    colorlinks,
    linkcolor={black},
    citecolor={blue!50!black},
    urlcolor={black}
}

\title{The Role of Flexible Connection in Accelerating Load Interconnection in Distribution Networks}

\author{Nan Gu \\
 Purdue University \\
 {\underline{ gu382@purdue.edu}} \And
 Ge Chen \\
 Purdue University \\
 {\underline{ chen4911@purdue.edu}} \And
 Junjie Qin \\
 Purdue University \\
 {\underline{ jq@purdue.edu}}}
\date{\empty}

\begin{document}
\addtolength{\abovedisplayskip}{-.05in}
\addtolength{\belowdisplayskip}{-.035in}
\addtolength{\belowcaptionskip}{-.015in}
\maketitle
\begin{abstract}
    \vspace{-1pt}
This paper investigates the role of flexible connection in accelerating the interconnection of large loads amid rising electricity demand from data centers and electrification. 
Flexible connection allows new loads to defer or curtail consumption during rare, grid-constrained periods, enabling faster access without major infrastructure upgrades. 
To quantify how flexible connection unlocks load hosting capacity, we formulate a flexibility-aware hosting capacity analysis problem that explicitly limits the number of utility-controlled interventions per year, ensuring infrequent disruption. 
Efficient solution methods are developed for this nonconvex problem and applied to real load data and test feeders. 
Empirical results reveal that modest flexibility, i.e., few interventions with small curtailments or delays, can unlock substantial hosting capacity. 
Theoretical analysis further explains and generalizes these findings, highlighting the broad potential of flexible connection.
\end{abstract}

\subsubsection*{Keywords:}

Flexible Connection,  Hosting Capacity Analysis, Distribution Grids, Interconnection Queues

\vspace{-2pt}
\section{Introduction}\vspace{-2mm}

The electric power sector is entering a new era of load growth, one that is expected to far exceed historical trends and is driven by a confluence of transformative forces. At the forefront is the surge in electricity demand from artificial intelligence data centers, whose compute-intensive workloads are expected to more than double global data center consumption to over 945 TWh by 2030 \cite{energy_ai_iea}. 
Simultaneously, the electrification of transportation and buildings is accelerating: Electric vehicles (EVs) may contribute 500-800 TWh of annual demand by 2030, while heat pump adoption is projected to add several hundred additional TWh \cite{outlook_demand_iea, outlook_2024_iea}.
Yet, the expansion of grid infrastructure lags behind. Lengthy interconnection queues \cite{Lawrence} and physical grid constraints make it increasingly clear that the current approach of sizing infrastructure for inflexible peak loads can hardly scale with the pace of electrification.

This challenge of serving a rapidly growing, diverse pool of users within finite infrastructure capacity is not unique to power systems. Other shared infrastructure sectors, such as communication networks and road transportation systems, have long grappled with similar pressures and responded by adopting \emph{over-subscription} as a design principle. Rather than provisioning for the absolute peak demand of every user, these systems assume that not all users will require maximum capacity simultaneously, and that when demand does surge, it can often be managed by flexibly shifting or throttling usage without compromising overall service. Internet service providers, for example, routinely connect far more users than their backbone bandwidth can support at peak, relying on statistical multiplexing and buffering to ensure reasonable performance. Similarly, urban road networks regularly operate above their nominal capacity during rush hours, absorbing excess demand through congestion and delays. In both cases, demand is allowed to exceed infrastructure limits at times, with the understanding that the resulting performance degradation, i.e., slower data speeds or longer travel times, is acceptable. In contrast, electric power systems are deliberately \emph{over-provisioned rather than oversubscribed}, prioritizing real-time balance and system security over resource utilization. This design principle is driven by the fact that, unlike data packets and cars, electrons cannot wait. 

However, not all electricity services require instantaneous, uninterrupted delivery. A growing class of loads can tolerate delays, curtailment, or rescheduling without degrading their core function. This  flexibility enables a new interconnection paradigm: \emph{flexible connection}.
Rather than waiting for costly and time-consuming grid capacity upgrades, new loads can be interconnected under flexible service agreements that explicitly account for their ability to operate within dynamic grid constraints. In the event of grid stress, controllable loads, such as EV chargers and deferrable computing tasks in data centers, can be reliably curtailed or shifted in real time, enabling safe oversubscription of grid capacity without risking overloads.

Beyond accelerating load interconnection, in the operational time frame, flexible connection differs fundamentally from existing grid management practices in both intent and implementation.
Unlike traditional demand response, which often relies on voluntary customer participation or price signals, flexible connection involves pre-arranged, utility-enforceable control of load to ensure predictable system performance. Unlike rolling blackouts, which indiscriminately cut power under emergency conditions, flexible connection enables targeted, non-disruptive adjustments to selected loads, preserving both grid reliability and service continuity. Importantly, many of these loads would otherwise face long delays and high costs to secure interconnection under conventional processes. With flexible connection, they can be brought online much sooner, avoid or defer major infrastructure investments, and in many cases experience only infrequent curtailment events over the course of a year, making it a highly attractive option for both customers and grid operators. 
This approach is already gaining traction; see pilots conducted by major utilities such as PG\&E \cite{PGandE} and ComEd \cite{doeflexCon}. 


Despite the great potential in flexible connection, critical questions remain open: 
\emph{How to efficiently quantify the hosting capacity of distribution grids, i.e., the maximum new load that can be safely interconnected, accounting for demand flexibility?}
\emph{Can substantial grid  capacities be unlocked by infrequent interventions to loads?} \emph{If so, why?}
Answering these questions is the key to {a} wider adoption of such programs to address  load interconnection challenges faced by utilities. 

To answer these questions, we start by formulating the problem of flexibility-aware hosting capacity analysis (FA-HCA) in Section~\ref{sec:problem}, where we incorporate two flexibility models, i.e., curtailment flexibility (CF) and delay flexibility (DF), and distribution network constraints. 
 We then propose efficient methods for solving the FA-HCA problem, empirically evaluate the hosting capacity unlocked by infrequent interventions, and develop theoretical justifications of the key empirical observations. These are done for copperplate network with CF in Section~\ref{sec:CM} and DF in Section~\ref{sec:DM}, and then extended to general radial distribution grids in Section~\ref{sec:network}. 
In this process, we make the following key contributions:
\begin{enumerate}[label={\alph*)}]
	\item By explicitly limiting the maximum number of interventions throughout the planning horizon, our FA-HCA formulation ensures the load connected through flexible connection is uninterrupted for most of the time. The resulting nonconvex optimization is then solved efficiently utilizing  the order statistics of \emph{dynamic hosting capacity}. 
	\item We establish key empirical observations ({\bf KO1-KO4} in the paper) on the hosting capacity that can be unlocked by flexible connection. Chiefly, infrequent interventions can unlock significant hosting capacities. Most of the interventions require {a} small depth of curtailment for CF, or delay window length for DF. These also hold for the network case.
	\item We develop theoretical models and results to justify these empirical insights, including a probabilistic model linking the unlocked hosting capacity with the tail distribution of the aggregate load, a formal connection between hosting capacities unlocked by CF and DF, and a novel structural equivalence between the copperplate and network cases. 
These results in turn imply the empirical observations likely hold broadly beyond the tested datasets.
\end{enumerate}

\noindent{\emph{Related Literature:}} 
{Early hosting capacity analysis focuses on the generation side, prioritizing renewable resources \cite{10040563} and their flexible interconnection \cite{9583062, EPRI}.}
The rising presence of EVs \cite{9638170}, data centers \cite{10.1145/3632775.3661959}, and building electrification \cite{elmallah2022can} has shifted focus toward load-side hosting capacity, often studied through scenario-based simulations \cite{9638170} and stochastic optimization \cite{10689225} using inflexible load profiles that yield conservative estimates.
Meanwhile, the flexibility potential of emerging loads \cite{shao2012grid, hao2014aggregate, 7039172} has been widely recognized, though its targeted use for improving hosting capacity remains under-explored.
Recent efforts have explored optimization-based strategies with device coordination \cite{kamruzzaman2020reliability, rana2022ev,almutairi2024hierarchical}, but often rely on complex, data-intensive interventions that limit practical implementation.
One recent study \cite{norris2025rethinking} shares the same high-level idea that minimal interventions to loads can substantially enhance hosting capacity.  It focuses on evaluating how infrequent curtailments can increase {the load} served by U.S. transmission systems. 
In contrast, we consider both curtailment and delay-based interventions, and address distribution network constraints.
\looseness = -2

\section{Formulation}\vspace{-1mm}
\label{sec:problem}
\vspace{-1mm}


Consider a single-phase, radial distribution feeder.
Suppose that the network has $n+1$ buses, with the substation/root bus labelled as bus $0$.
In this paper, we will perform our hosting capacity analysis focusing on a planning horizon with a finite set $\mathcal T$ of discrete time intervals, where the number of time slots is denoted by $T$. 
Each time interval can be a metering interval, which usually spans 15 minutes or 30 minutes for US utilities. 

\def\inew{i^\dagger}
\def\lnew{\ell^\dagger}
\def\blnew{\bm{\ell}^\dagger}
\def\lbnew{\hat{\ell}^\dagger}
\def\blbnew{\bm{\hat{\ell}}^\dagger}
\def\cnew{C}
\def\lag{\ell^\mathrm{agg}}
\def\cs{C^\star}
\def\cres{C^\mathrm{res}}
Denote the existing real power load at bus $i=1, \dots, n$ in period $t\in \mathcal T$ by $\ell_i(t)$. This information may come from available historical data. We are interested in how much capacity of certain new loads may be connected at a bus $\inew$ under flexible connection. This new load may represent a large EV charging station, a data center, or an aggregation of smaller loads to be connected to the distribution network through the bus.  For the new load, the real power consumption is calculated by 
\begin{equation}
	\lnew(t) = \cnew \lbnew(t),    \quad t \in \mathcal T,
	\vspace{-1pt}
\end{equation}
where $\cnew$ is the real-power capacity of the new load and $\blbnew \in \mathbb R^T$ is the normalized time-varying profile for the new load such that $0 \le \lbnew(t) \le 1$ for all $t\in \mathcal T$. 

In order to characterize the maximum $\cnew$ (i.e., the load hosting capacity at bus $\inew$) that the distribution feeder can support under the flexible connection program without violating  network constraints, we next outline our models for demand flexibility and power flow. 

\subsection{Flexibility Models}\vspace{-1mm}
We focus on two modes of demand flexibility for the new load interconnected under the flexible connection program: \emph{curtailment} and \emph{delay}. 
The curtailment flexibility  model is conceptually simple, whereas the delay flexibility  model accurately characterizes flexibility from loads such as EVs. 
For both flexibility models, we aim to have \emph{very infrequent interventions} to the new load, either directly implemented by the utility  or through a third-party service provider such as an aggregator. To this end, we explicitly limit the number of interventions over the planning horizon (e.g., a year). For both models, we denote the vector of load modification by $\mathbf u \in \mathbb R^T$ so the modified load will be $ \blnew+ \mathbf u$.


\emph{a) Curtailment Flexibility (CF):} The new load agrees to be curtailed in up to $K$ time slots over the planning horizon $\mathcal T$. 
With CF, the modification vector must satisfy
\begin{equation}\label{eq:ucm}
	\mathbf u \in \mathcal U^\mathrm{CF}_K:= \{\mathbf u\in \mathbb R^T: \mathbf u \le \mathbf 0 \quad \mbox{and} \quad \|\mathbf{u}\|_0 \le K\}, 
\end{equation}
where the $0$-``norm'' of a vector, i.e., $\|\cdot\|_0$, returns the number of non-zero elements of the vector. 

\emph{b) Delay Flexibility (DF):} The new load agrees to experience up to $K$ delay events over the planning horizon $\mathcal T$; in each delay event, the load in the intervened time slot may be delayed for at most $D$ time slots. Motivated by EV charging flexibility, we allow \emph{fractional delay} in the sense that if time $t$ is picked for delay intervention, a portion of the new load's power consumption in slot $t$ can be shifted towards the next $D$ time slots. 
A feasible load modification vector $\mathbf u$ is thus determined by two factors: \emph{when to intervene} and \emph{how to shift} the loads during each intervention. For the former, we use binary vector $\mathbf x_k \in \mathbb R^T$ to embed the picked time for the $k$-th intervention: $x_k(t) = 1$ if the $k$-th intervention occurs at time $t$ and $0$ otherwise for $k=1, \dots, K$. These vectors need to satisfy:
\begin{subequations}\label{eq:xconst}
	\begin{align}
	&x_k(t) \in \{0,1\}, \quad k=1, \dots, K,\ t \in \mathcal T, \label{eq:xconst1}\\
	&x_k(t) = 0,\! \qquad\quad k = 1, \dots, K,\ t > T-D,  \label{eq:xconst2}\\
	&\mathbf 1^\top \mathbf x_k \le  1, \qquad\quad k = 1, \dots, K,\label{eq:xconst3}
	\end{align}
\end{subequations} 
where \eqref{eq:xconst2} ensures that all delayed load can be fully served within the planning horizon, and \eqref{eq:xconst3} enforces that at most $K$ delay events are scheduled. To characterize how to shift load in each delay event, let  $\mathbf U \in \mathbb R^{ (D+1) \times K}$ be defined such that the $k$-th column of $\mathbf U$, denoted by $\mathbf U_k \in \mathbb R^{D+1}$, contains the load modification associated with the $k$-th delay event, with $U_{k,1}\le 0$ representing the load reduction at time  $t$, and $U_{k,\tau}\ge 0$ for $\tau =2, \dots, D+1$ representing the load increase in subsequent $D$ slots. It follows that matrix $\mathbf U$ must satisfy the following constraints, for $k = 1, \dots, K$ and  $\tau =2, \dots, D+1,$
\begin{equation}\label{eq:Uconst}
	U_{k,1} \le 0,\quad U_{k,\tau} \ge 0, \quad \mathbf 1^\top \mathbf U_k =0,
\end{equation}
where the last constraint ensures that all delayed energy consumption is served within the next $D$ time slots.


It remains to translate individual delay events' impact to the aggregate load modification vector $\mathbf u$. To this end, we wish to create an extended version of $\mathbf U_k\in \mathbb R^{D+1}$, denoted by $\widetilde{\mathbf U}_k \in \mathbb R^T$ for each $k$, such that $\widetilde{\mathbf U}_k$ contains elements of $\mathbf U_k$ in its $t_k$-th to $(t_k+D+1)$-th positions and $0$ elsewhere, if $x_k(t_k)=1$. Specifically, $\widetilde{\mathbf U}_k$ can be calculated by convolving the binary intervention signal $\mathbf x_k$ with  $\mathbf U_k$ to distribute the values in $\mathbf U_k$ over time: \vspace{-.1in}
\begin{equation}
	\widetilde{ U}_{k,t}(\mathbf U_k, \mathbf x_k) := \sum_{d=0}^{D} U_{k,d+1}x_k(t-d), \quad t \in \mathcal T, \vspace{-.1in}
\end{equation}
where $x_k(\tau):=0$ for $\tau \le 0$. Then we have \vspace{-.1in}
\begin{equation}
\label{eq:DM-u}
	\mathbf u = \sum_{k=1}^K \widetilde{\mathbf U}_{k}(\mathbf U_k, \mathbf x_k). 
	\vspace{-.1in}
\end{equation}

 We can summarize the constraints for the DF case as
 \(
 	\mathbf u \in \mathcal U^\mathrm{DF}_K, 
 \)
 where  set $\mathcal U^\mathrm{DF}_K$ contains all vectors in the form of \eqref{eq:DM-u}, with $\mathbf U$ satisfying \eqref{eq:Uconst} and $\mathbf x$ satisfying \eqref{eq:xconst}. 
%
\vspace{-1mm}
\subsection{Power Flow Constraints}\vspace{-1mm}
The power flow induced by the existing load and the new load, potentially modified due to infrequent interventions as allowed by the flexible connection program,
must satisfy the physical constraints of the network. 
Let the vector of real power injection over the network other than the substation bus $0$ at time $t$ be denoted by $\mathbf p(t) \in \mathbb R^{n}$, which takes the form of 
\begin{equation}\label{eq:p}
	\mathbf p(t) = \mathbf g(t) - \bm \ell(t) - [\lnew(t)+u(t)]\mathbf e_{\inew} ,
\end{equation}
where $\mathbf e_{\inew} \in \mathbb R^{n}$ denotes the elementary vector whose $\inew$-th element is 1 and all other elements are 0, and $ g_i(t)$ denotes 
the real power generation from distributed solar panels. 
For our purpose of determining the load hosting capacity and as we cannot count on solar production to ensure grid reliability when the sun is not shining, we set $\mathbf g(t) \equiv \mathbf 0$ for all $t$ as a conservative modeling assumption.  
Let $\mathbf q(t) \in \mathbb R^{n}$ be the vector of reactive power injection over the network except the bus 0. Denote the ratio of reactive to real power by $\blue{\bm \eta(t)}\in \mathbb R^{n}$, which can be computed using power factor information  from historical data. Then the reactive power injection can be written as 
\vspace{-10pt}
\begin{equation}\label{eq:q}
	\mathbf q(t) = \mathrm{diag}(\blue{\bm \eta(t)})  \mathbf p(t). 
	\vspace{-2pt}
\end{equation}

In general,  distribution network constraints can be summarized as a feasible complex power injection region, which can be equivalently modeled by a feasible real power injection region $\mathcal P \subset \mathbb R^n$ here as the reactive power and real power are related via \eqref{eq:q}. Thus a load vector is feasible given distribution grid constraints if 
\vspace{-2pt}
\begin{equation}
	\bm \ell(t) + [\lnew(t)+u(t)]\mathbf e_{\inew} \in \mathcal L,
	\vspace{-2pt}
\end{equation}
where $\mathcal L \subset \mathrm R^{n}$, denotes the set of feasible loads given the network constraints, contains loads such that the corresponding $\mathbf p(t)\in \mathcal P$. 


More concretely and to facilitate analytical results in this paper, we adopt the standard linearized DistFlow model and consider the following network constraints:

\emph{a) Substation Transformer Capacity Constraint}, converted to the corresponding real power limit given power factor information, takes the form of 
\vspace{-2pt}
\begin{equation}\label{eq:poolcon}
	p_0(t) \le \overline{p}_0, \quad p_0(t) + \mathbf 1^\top \mathbf p(t) = 0, \quad t \in \mathcal T, 
	\vspace{-2pt}
\end{equation}
where $p_0(t)$ denotes the real power flowing through the substation bus (i.e., the point of common coupling) from the upstream network to this distribution feeder and $\overline p_0$ is  its upper bound. Since  the network is lossless under linearized DistFlow assumptions, $p_0(t) \ge 0$ simply collects the total load across the network at time $t$. 

\emph{b) Voltage Constraints} can be written as
\begin{subequations}\label{eq:vcon}
	\begin{align}
		\mathbf v(t) = v_0 \mathbf 1 + \mathbf R \mathbf p(t) + \mathbf X \mathbf q(t), \quad t \in \mathcal T, \\
		\underline{\mathbf{v}} \le \mathbf v(t) \le \overline{\mathbf{v}}, \quad t \in \mathcal T, 
	\end{align}
\end{subequations}
where 
$\mathbf v(t) \in \mathbb R^n$ denotes the voltage magnitude for all buses except the substation bus; $v_0$ denotes the voltage magnitude at the substation bus which is usually 1.0 with per unit analysis; matrices $(\mathbf R, \mathbf X)$ depend on the feeder topology and line impedances; and $\underline{\mathbf{v}}$ and $\overline{\mathbf{v}}$ are the known bounds for voltage magnitudes. 

Equipped with constraints \eqref{eq:poolcon} and \eqref{eq:vcon} and together with \eqref{eq:p} and \eqref{eq:q}, we can define the feasible injection region $\mathcal P$ and feasible set of loads $\mathcal L$ accordingly. 

\subsection{Flexibility-Aware Hosting Capacity Analysis}\vspace{-1mm}
Collecting our flexibility models and grid constraints, we arrive at the following formulation of \emph{flexibility-aware hosting capacity analysis} (FA-HCA): \vspace{-.05in}
\begin{subequations}\label{opt:hca}
	\begin{align}
		\max_{\cnew, \mathbf u} \quad & \cnew \vspace{-.05in}\\
		\mbox{s.t.} \ \quad & \lnew(t) = \cnew \lbnew(t), \quad\ t \in \mathcal T, \label{opt:hca:c1}\\
		&\bm \ell(t) +[\lnew(t)+u(t)]\mathbf e_{\inew} \in \mathcal L,\quad t \in \mathcal T, \label{opt:hca:c2}\\
		& \mathbf u \in \mathcal U_K, \label{opt:hca:c3}
		 \vspace{-.2in}
	\end{align}
\end{subequations}
where $\mathcal U_K$ is either $\mathcal U^\mathrm{CF}_K$ 
or $\mathcal U^\mathrm{DF}_K$
, depending on which flexibility model is used. 
Denote the optimal value of this optimization, given intervention budget $K$, by $\cs_K$. We also use $\cs_0$ to denote the hosting capacity without considering flexibility. Problem~\eqref{opt:hca} in its current form is nonconvex because of the 
\blue{nonconvexity} of set 
$\mathcal U_K$, which arises due to~\eqref{eq:ucm} for CF and~\eqref{eq:xconst1} and \eqref{eq:DM-u} for DF. 

In  remaining sections,  starting from the copperplate case and then for general networks, we  develop efficient methods to solve~\eqref{opt:hca},  obtain key empirical observations on flexible connection's role  for unlocking hosting capacities in distribution grids, and establish {a} theoretical understanding of such empirical observations. 


\emph{Proofs of all theoretical results can be found in the extended version of the paper \cite{extended_version}.}


\section{Copperplate Case with Curtailment Flexibility}	\vspace{-1mm}
\label{sec:CM}
We begin with a stylized setting where the distribution network is modeled as a copperplate: the transformer constraint~\eqref{eq:poolcon} is enforced, while voltage constraints~\eqref{eq:vcon} are omitted. This allows us to isolate the interplay among the temporal characteristics of the existing load, the flexibility of the new load, and the transformer capacity limit. Since load locations are irrelevant under the copperplate model, we denote the aggregate existing load as
\begin{equation}
	\lag (t):= \mathbf 1^\top \bm \ell(t), \quad t \in \mathcal T.
\end{equation}
We will treat the curtailment flexibility (CF)   in this section, with the delay flexibility (DF)   handled later. 

\subsection{Methods for FA-HCA}\label{sec:HCA:CM}\vspace{-1mm}
The FA-HCA problem with CF and copperplate network, despite still nonconvex due to the constraint $\|\mathbf u\|_0 \le K$, can be solved efficiently leveraging the following key quantity. 
\begin{definition}[Dynamic Hosting Capacity and Order Statistics]\label{def:dhc}
	For each $t\in \mathcal T$, we refer to 
	\begin{equation}
		\cres(t):= \overline p_0 - \lag(t) \ \mbox{and}\  \cnew(t):= \cres(t)/\lbnew(t)
	\end{equation}
	as the dynamic residual capacity and dynamic hosting capacity for time $t$, respectively. 
	Further, let the $s$-th (lower) order statistics of $\{\cnew(t): t \in \mathcal T\}$ be $\cnew[s]$, and let the time index corresponding to the $s$-th order statistics be $t_s$. Mathematically, 
\begin{subequations}\label{eq:rtC}
\begin{align}
	&\cnew[s] = \cnew(t_s), \,\, \qquad s \in \mathcal T, \\
	&\cnew[s] \le \cnew[s+1], \quad s \in \mathcal T \backslash \{T\}.\vspace{-.05in}
\end{align}
\end{subequations}
\end{definition}

The dynamic hosting capacity characterizes how much new load capacity  can be accommodated   at time $t$. Using this notion, constraint~\eqref{opt:hca:c2} takes the form of 
\begin{equation}\label{eq:copperplate:cons}
	C \le C(t) - (u(t)/\lbnew(t)), \quad t \in \mathcal T. 
\end{equation}
We can then solve~\eqref{opt:hca} analytically as follows:

\begin{proposition}[Solving FA-HCA: Copperplate  with CF]\label{prop:HCAsoln:CM}
	The optimal value of \eqref{opt:hca} is
	\begin{equation}
		\cs_K = C[K+1].\vspace{-.05in}
	\end{equation}
	 The optimal load modification is \vspace{-.05in}
	\begin{equation}\label{eq:ustar:ccm}
		u^\star(t) = \begin{cases}
			 \cres(t) - \cs_K \lbnew(t), & \mbox{if } t = t_s,\,\, s\le K,\\ 
			0, & \mbox{ otherwise.}
		\end{cases}\vspace{-.05in}
	\end{equation}
\end{proposition}

Proposition~\ref{prop:HCAsoln:CM} suggests  \eqref{opt:hca} can be solved by sorting  $\cnew(t)$ values which determine the hosting capacity that can be supported at time $t$. Given the intervention budget $K$, the $K$ critical  slots with the lowest $\cnew(t)$ values will be curtailed. As a result, the achievable hosting capacity accounting for CF is the $(K+1)$-th lowest dynamic hosting capacity, and the minimum curtailed energy in the $K$ time slots {is} sufficient to bring up the dynamic hosting capacity in these slots to $\cnew[K+1]$ per~\eqref{eq:ustar:ccm}.

\vspace{-0.8mm}
\subsection{Empirical Observations}\label{sec:EO:CM}\vspace{-1.2mm}
Equipped with Proposition~\ref{prop:HCAsoln:CM}, we compute the flexibility-aware hosting capacity under the copperplate model with CF. To gain empirical insights, we apply the result to loads from the NREL U.S. ResStock dataset~\cite{osti_1854582}. The  existing load time series $\{\lag(t)\!:\! t\! \in\! \mathcal T\}$ is constructed by aggregating 15-minute power consumption data from 500 randomly selected buildings in Los Angeles County, California, for the full year of 2018 ($T\! =\! 35040$). We consider a scenario in which an apartment developer plans to retrofit buildings with a large number of level-2 chargers, using apartment EV charging profiles from~\cite{sorensen2021analysis} to generate the normalized new load $\{\lbnew(t):\! t\! \in \!\mathcal T\}$. The existing load has a peak of 1.53~MW, and we assume 10\% residual transformer capacity {(reflecting limited headroom)}, yielding $\overline p_0\! =\! 1.683$~MW. Our empirical analysis reveals the following key observations (KOs):
\begin{enumerate}[label=\bfseries KO\arabic*., ref=KO\arabic*]
	\item [\bf KO1.] Infrequent interventions unlock significant hosting capacities.
	\item [\bf KO2.] Most interventions require small curtailments. 
\end{enumerate}

\begin{figure}[h]
\vspace{-.08in}
    \centering
    \includegraphics[width=\linewidth]{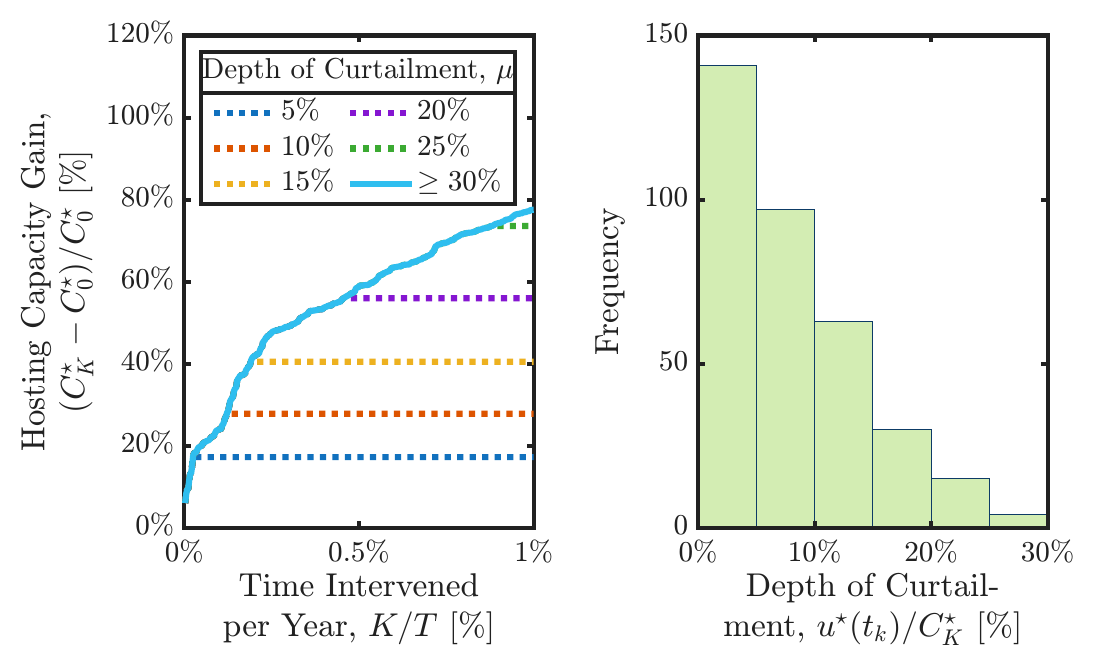}
	\vspace{-.15in}
    \caption{ Hosting capacity gain for CF (left) and  depth of curtailment requirement distribution with $K/T=1\%$ (right). }
    \vspace{-.08in}
    \label{fig:curtail_info:CM}
\end{figure}

The left panel of Fig.~\ref{fig:curtail_info:CM} illustrates {\bf KO1}, showing how hosting capacity gain varies with the percentage of time intervened per year  and the bound on curtailment depth. The gain is measured relative to the baseline hosting capacity without flexibility. The intervention budget $K$ in~\eqref{opt:hca} is expressed as $(K/T) \times 100\%$ on the x-axis, focusing on small $K$ values—up to 1\% of annual time slots. The light blue solid line demonstrates a 77.55\% capacity gain with just 1\% of time intervened. In absolute terms, the hosting capacity increases from 0.88~MW (baseline) to 1.56~MW with flexibility. The gain beyond the 0.15~MW residual capacity (i.e., $1.53 \times 10\%$) arises from the fact that $\lbnew(t) < 1$ during critical time slots when curtailment is applied. To assess the power reduction required per curtailment event, we introduce a bound on curtailment depth $\mu$ and solve~\eqref{opt:hca} with the added constraint $|u(t)| \le \mu \cnew$ for all $t \in \mathcal T$. The results, shown as dashed lines in the left panel of Fig.~\ref{fig:curtail_info:CM}, coincide with the solid curve until they plateau. These lines show that a large share of the hosting capacity gain can be achieved even under a strict, uniform curtailment limit. For example, a 20\% curtailment bound yields nearly 60\% capacity gain, while a 30\% bound suffices to capture the full benefit of curtailment-based flexibility.

Instead of imposing a uniform curtailment bound, we can leave curtailment unconstrained and use~\eqref{eq:ustar:ccm} to compute the required curtailment at each intervention time. The resulting distribution, shown in the right panel of Fig.~\ref{fig:curtail_info:CM} for $K = 350$ (about 1\% of the year), shows that most interventions require only modest curtailment. Notably, curtailment exceeding 25\% occurs in just 4 time slots, supporting {\bf KO2}.


\subsection{Probabilistic Modeling and Universality}\label{sec:theory:CM}\vspace{-1mm}
While results presented in the previous subsection are derived from a particular set of loads, further tests with other loads suggest that {\bf KO1} and {\bf KO2} persist across diverse settings. 
In this subsection, we aim to develop a probabilistic model to explain such empirical observations, which will in turn imply that these observations are likely to occur broadly for typical systems and load profiles.
To this end, we impose the following assumptions to facilitate analysis:
\def\FHL{F^\mathrm{HL}}
\def\fHL{f^\mathrm{HL}}
\def\lagbdd{\overline{L}}
\begin{enumerate}[label=\bfseries A\arabic*., ref=A\arabic*]
	\item [\bf A1.] $\lbnew(t)=1$ for all $t$. 
	\item [\bf A2.] The high values of the aggregate existing load, i.e., $\{\lag(t): \lag(t) \ge L, \,  t \in \mathcal T\}$ for some threshold $L >0$, are independent and identically distributed (i.i.d.) random variables whose distribution has cumulative distribution function (CDF) $\FHL(\cdot)$ and probabilistic density function (PDF) $\fHL(\cdot)$. The support of $\fHL$ is a closed interval of $\mathbb R$ with its right endpoint being $ \lagbdd < \infty$. 
\end{enumerate} 

With {\bf A1}, we adopt a conservative model for the new load, and thus can focus on analyzing how the existing load profiles impact the flexibility-aware hosting capacity.  In this analysis, it turns out that a critical determinant of the hosting capacity is the \emph{right tail} of the empirical distribution of $\{\lag(t): t\in \mathcal T\}$. {\bf A2} introduces a tractable model for this right tail, with $L$ determining the cutoff value, and $\FHL$ or $\fHL$ characterizing its shape. Given an $L$ value, the number of time slots with the existing load greater than or equal to $L$ is denoted as $T_L:=\beta_L T$ with $\beta_L\in [0,1]$. For instance, if $L$ is selected to be the median of the existing load process, then $\beta_L = 0.5$. We also impose a finite upper bound on the existing load, i.e., $\lagbdd$, which may be derived from the sum of max power values associated with contracted electric service levels for all users.

Under these assumptions, the hosting capacities, with and without considering the flexibility, i.e., $\cs_K$ and $\cs_0$,  are both random variables. We have the following results on their distribution and expected values.

\begin{theorem}[Distribution of $\cs_K$]\label{thm:1}
	Under {\bf A1} and {\bf A2}, for $K=0,1,\dots$, the following holds for $\cs_K$.
	\begin{enumerate}[label={\alph*)}]
		\item For any $c>0$, with $\rho_c:= 1-\FHL(\overline p_0 - c)$, 
		\begin{equation}
			\mathbb P(\cs_K > c) = \sum_{k=0}^K {T_L \choose k} \rho_c^k (1-\rho_c)^{T_L-k}.
			\label{thm1-a}
		\end{equation}
		\item For $c>0$ and $T_L$ large, with $\lambda_c:= T_L \rho_c$, we have 
		\begin{equation}\label{eq:cs:cdf:ap}
		\!\!\!\!\!\!\!\!\!\!\!\!\!\!\mathbb P(\cs_K > c)\! \approx \!\sum_{k=0}^K \frac{\lambda_c^k}{k!}\exp(-\lambda_c)	= \mathbb P (\mathrm{Pois}(\lambda_c) \le K),
		\end{equation} 
		\begin{equation}
			\mathbb E [\cs_K] \approx\overline p_0 - (\FHL)^{-1}\left(1- \frac{K+1}{T_L+1}\right),
			\label{eq:thm1:expectation}
		\end{equation} 
		where $\mathrm{Pois}(\lambda_c)$ denotes a Poisson random variable with rate parameter $\lambda_c$. 			
	\end{enumerate}
\end{theorem}

Theorem~\ref{thm:1} establishes the distribution of $\cs_K$ leveraging its connection to the (lower) order statistics $\cnew[K+1]$ and the corresponding (upper) order statistics of the existing load. The approximate expression~\eqref{eq:cs:cdf:ap} is easier to interpret: For any target hosting capacity $c>0$, the probability that $\cs_K>c$ is the same as that a Poisson random variable with {a} certain rate parameter being no greater than $K$. The
rate here is the probability of the aggregate load going beyond $\overline p_0 -c$, i.e., $\rho_c$, scaled by the number of high-load periods, i.e., $T_L$; thus the event associated with the Poisson random variable precisely models situations where up to $K$ curtailments can support a hosting capacity value $c$. The approximation to the expected value of $\cs_K$ is derived from the well-known expected value formula for order statistics of uniform random variables and Taylor expansion \cite[80]{david2004order}. 

To further understand how $\cs_K$ depends on $K$ and given the critical role played by the (upper) order statistics of the aggregate load in Theorem~\ref{thm:1}, we  draw inspirations from \emph{extreme value theory} which  characterizes the distributions of extreme values of stochastic processes. In particular, for extreme values of bounded stochastic processes, the Weibull  extreme value distributions are commonly used. This motivates us to adopt the following family of distributions \cite[38]{mikosch2024extreme} to model the right tail of the aggregate  load:
\begin{enumerate}[label=\bfseries A\arabic*., ref=A\arabic*]
\setcounter{enumi}{2}
\item [\bf A3.]  The right tail of the aggregate load distribution, i.e.,  $\fHL(x)$,   decays near the right endpoint as 
	\begin{equation}
\fHL(x) = \kappa \left[ 1 - (x/\lagbdd)\right]^\alpha, \quad L \le x\le \lagbdd.
\label{eq:tail}
\end{equation}
where $\kappa >0$ and $\alpha>0$ are constant  parameters.  
\end{enumerate}

\begin{figure}[h]
\vspace{-.1in}
    \centering
    \includegraphics[width=\linewidth]{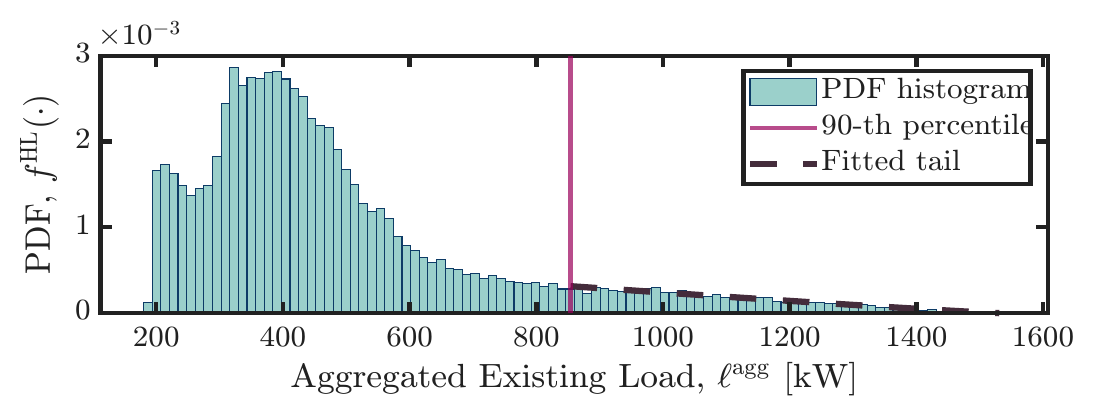}
    \vspace{-.2in}
    \caption{Distribution of the existing load.}
    \vspace{-.1in}
    \label{fig:distribution_old}
\end{figure}

Given the aggregate load data and a pre-determined cutoff value, we can identify the parameters $(\kappa,\alpha)$ by fitting {them} to the empirical distribution of the right tail. Fig.~\ref{fig:distribution_old} depicts such a fit when $L$ is set to the 90-th percentile of the aggregate load, and $\lagbdd$ is set to the maximum observed aggregate load. For this example, $(\kappa, \alpha) = (7.71\times 10^{-3}, 1.10)$.  

Focusing on the expected hosting capacity values, we have the following results. 
\begin{theorem}[Marginal Gain \& Depth of Curtailment]\label{thm:2}
	Under {\bf A1}, {\bf A2}, and {\bf A3}, for any $K=0, 1, \dots$, we have: 
	\begin{enumerate}[label={\alph*)}]
	\item The expected hosting capacity is approximated as \label{thm2-a}
	 \begin{equation}
    \mathbb E [\cs_K] \approx \overline p_0 - \lagbdd + (\lagbdd - L)\left(\frac{K+1}{T_L+1}\right)^{\frac{1}{\alpha+1}}.
    \label{eq:expected_fit}
    \end{equation}
    \item The marginal hosting capacity gain, defined as $g_{K+1}:= \mathbb E[\cs_{K+1}] - \mathbb E[\cs_K]$, is decreasing in $K$. 
    \item The set containing expected depth of curtailment requirements for the $K$ interventions, denoted by $\mathcal R_K:=\{ r_k:=- \mathbb E[u^\star(t_k)] >0: k=1, \dots, K\}$, concentrates around the lower values in the set. Mathematically, $r_k$ is the empirical $\gamma_k$-quantile of the set, where $\gamma_k = |\{r\in \mathcal R_K: r < r_k\}|/(K-1)$. We then have \label{thm2-c}
   \begin{equation}
    	 \frac{r_k- \min \mathcal R_K}{\max \mathcal R_K - \min \mathcal R_K}   < \gamma_k, \quad k =2, \dots, K-1,
    \label{eq: thm2-c}\vspace{-.05in}
    \end{equation}
    where $|\mathcal A|$ denotes the cardinality of any set $\mathcal A$. 
	\end{enumerate}

\end{theorem}

\begin{figure}[h]
\vspace{-.05in}
    \centering
    \includegraphics[width=\linewidth]{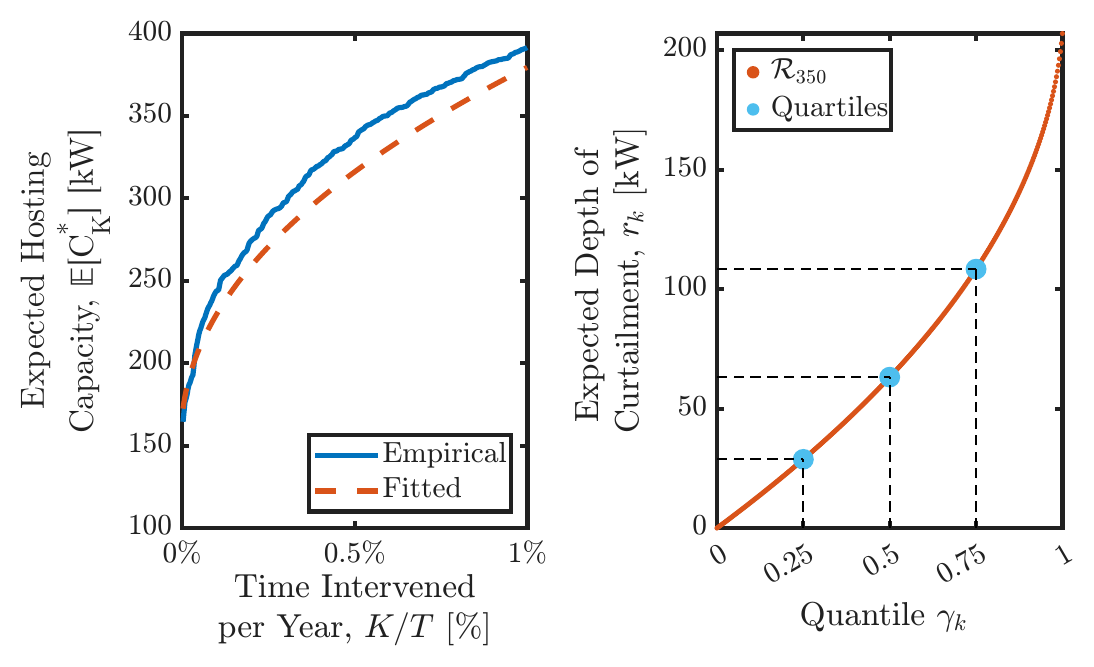}
	\vspace{-.1in}
    \caption{Expected  hosting capacity (left) and depth of curtailment distribution for $K=350$ (right).}
    \vspace{-.1in}
    \label{fig:thm2-results}
\end{figure}

With Weibull-type right tail, the expected hosting capacity can be expressed as a  function of $K$ per Theorem~\ref{thm:2}-\ref{thm2-a}, which is compared against the expected hosting capacity evaluated with \eqref{eq:thm1:expectation} and the empirical distribution of the aggregate load in the left panel of Fig.~\ref{fig:thm2-results}. 
A direct consequence of~\eqref{eq:expected_fit}, as also evident from the left panel of Fig.~\ref{fig:thm2-results}, is that $\mathbb E[\cs_K]$ is strictly concave in $K$. Thus when we consider increasing the intervention budget, the highest marginal gains (i.e., $g_K$ values) are obtained for small values of $K$, supporting {\bf KO1}. With a fixed $K$, Theorem~\ref{thm:2}-\ref{thm2-c} characterizes the distribution of the curtailment requirements. Among the curtailment requirements for the $K$ interventions in set $\mathcal R_K$, the 50-th percentile, for example, is less than the mid-point between the minimum and maximum curtailment requirement values (see  Fig.~\ref{fig:thm2-results}, right panel). This is consistent with {\bf KO2} as the majority of interventions have  relatively small depth of curtailment requirements. 

\vspace{-2mm}
\section{Copperplate Case with Delay Flexibility }\vspace{-2mm}
\label{sec:DM}
While conceptually simple, the curtailment model suffers from the pitfall that it does not account for the potential increase of load  after  curtailment interventions, i.e., the \emph{rebound effect}. 
In this section, by adopting the delay flexibility (DF) model, we explicitly consider such rebound effect. 
We continue to  focus on the transformer capacity constraint as in the previous section.

\subsection{Methods for FA-HCA} \vspace{-1mm}
The FA-HCA problem with DF is more complex than the CF case. 
Delay windows can overlap across interventions, introducing additional coupling between decisions. 
This coupling, combined with the nonconvexity from  integer variables $\mathbf{x}_k$'s and constraint~\eqref{eq:DM-u}, makes the problem challenging to solve.

To improve tractability, and motivated by empirical observations from the previous section, we assume that many time slots exhibit low aggregate existing load and thus sufficient dynamic residual capacity (cf. Definition~\ref{def:dhc}) to absorb shifted load from delay interventions. Let $\mathcal T_{k} := \{t \in \mathcal T : t_k \le t \le t_k + D\}$ denote the delay window for an intervention at time $t_k$, $k = 1, \dots, K$. If two delay windows overlap, i.e., $\mathcal T_k \cap \mathcal T_{k'} \neq \emptyset$ for $t_k < t_{k'}$, we merge them into a single window $\mathcal T_k \cup \mathcal T_{k'}$. After merging all overlapping windows, we obtain a finite set of disjoint delay windows, some of which may contain a \emph{cluster of interventions}. Let the $j$-th such window be $\mathcal T_j^{\mathrm{cl}}$, for $j = 1, \dots, J$. We now introduce a formal assumption to ensure each merged window contains at least one time slot with sufficiently large $\cres(t)$, in the following sense:
\begin{enumerate}[label=\bfseries A\arabic*., ref=A\arabic*]
\item [\bf A4.] Given $K$, $D$, and  $\mathcal T_j^\mathrm{cl}$ for $j=1, \dots,J,$
\vspace{-.05in}
\begin{equation}
\!\!\!\!	\sum_{t \in \mathcal T_{j}^\mathrm{cl}} \left[\cres(t) - \cnew[K] \lbnew(t) \right]\ge 0. 
\vspace{-.05in}
\end{equation}	
\end{enumerate}

Under {\bf A4}, FA-HCA with DF can be efficiently solved leveraging the sorted list of dynamic hosting capacity.

%
%

\begin{proposition}[Solving FA-HCA: Copperplate  with DF]\label{prop:HCAsoln:DM}
Suppose the times $t_k$, $k=1, \dots, K$, defined in Section~\ref{sec:HCA:CM}, satisfy $t_k < T-D$.
Under {\bf A4}, the solution of~\eqref{opt:hca} with DF and copperplate network, $(\mathbf u^\star, \cnew^\star_K)$, can be obtained from solving the following convex program:\vspace{-.05in}
\begin{subequations}    \label{opt:hda-sol}
    \begin{align}
    \max_{\cnew,\, \mathbf u,\, \mathbf U } \quad & \cnew \\
    \mathrm{s.t.}\quad\ &  \eqref{opt:hca:c1}, \eqref{eq:copperplate:cons}, \eqref{eq:Uconst}, \eqref{eq:DM-u}, \vspace{-.05in}
    \end{align}
\end{subequations}
with $\mathbf x_k$ \blue{in \eqref{eq:xconst}} determined as $x_k(t) = 1$ if $t = t_k$ and $x_k(t) =0$ otherwise.


%
%

\end{proposition}
\blue{
    Proposition~\ref{prop:HCAsoln:DM} implies that~\eqref{opt:hca} can be solved by first selecting the $K$ time slots with the smallest $\cnew(t)$ values for delay events. Their potentially overlapping delay windows are automatically coupled through~\eqref{eq:DM-u}, and the convex program~\eqref{opt:hda-sol} then co-optimizes the delayed energy and its allocation across all $K$ events.}

\begin{remark}[Dynamic Minimal Delay Requirement]\label{rk:Dk}
	For certain intervention time $t_k$'s, we may not need the entire $D$-slot window as the delayed load may be fully accommodated with fewer time slots. As such, among the solutions of \eqref{opt:hda-sol} given $\mathbf x_k$ values, we may identify  delay requirements $\{D_k: k = 1, \dots, K\}$ that are  minimal in the sense that we cannot further reduce any of the $D_k$'s without increasing another.  A heuristic algorithm is proposed in Appendix~E of \cite{extended_version}. 
\end{remark}
\vspace{-1mm}
\subsection{Empirical Observations}\vspace{-1mm}
As in Section~\ref{sec:EO:CM}, we evaluate the flexibility-aware hosting capacity  with DF. 
The key empirical observations that we obtain are summarized below, followed by discussions  supporting these observations. 
\begin{enumerate}[label=\bfseries KO\arabic*., ref=KO\arabic*]
\setcounter{enumi}{2}   
	\item DF unlocks the same hosting capacity as CF, provided a sufficiently long delay window $D$. 
	\item Most interventions require short delays. 
\end{enumerate}

\begin{figure}[h]
    \centering
    \includegraphics[width=\linewidth]{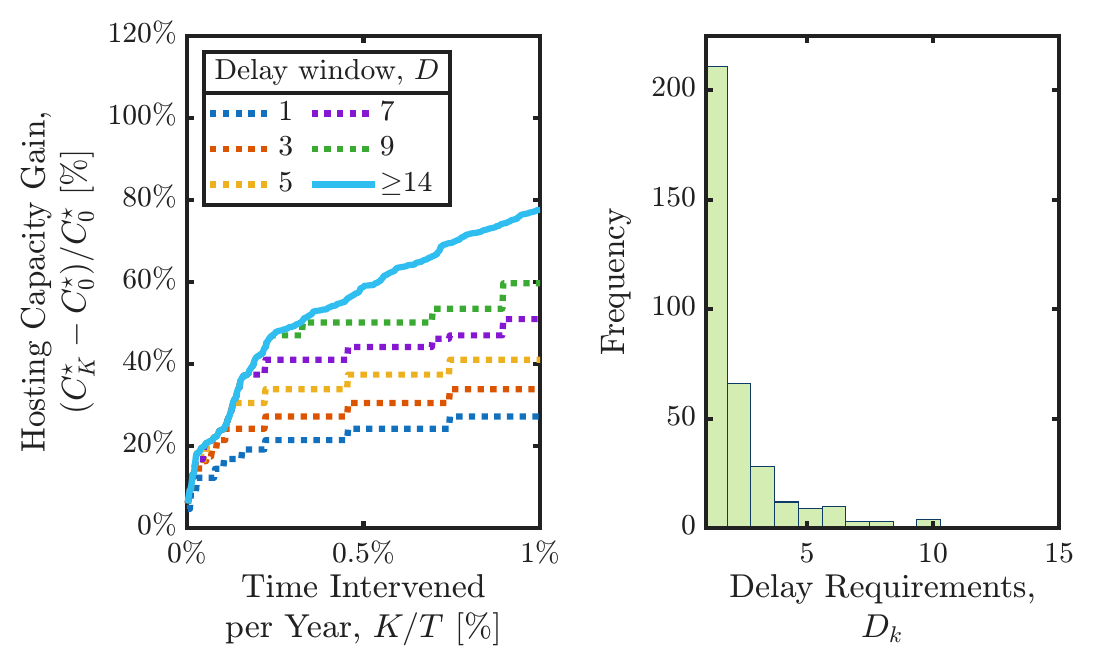}
    \caption{Hosting capacity gain for DF (left) and   delay requirement distribution with $K/T=1\%$ (right). }
    \label{fig:delay_info:DM}
    \vspace{-.1in}
\end{figure}

In the left panel of Fig.~\ref{fig:delay_info:DM}, the percentage hosting capacity gain, $(\cs_K-\cs_0)/\cs_0 \times 100\%$, is evaluated with different percentages of time intervened per year (i.e., $K/T$) and lengths of delay window $D$. {\bf KO3} is confirmed as the unlocked capacity from the CF with $D \ge 14$, i.e., 210 minutes, overlaps with that from the CF, for up to 1\% of time intervened in a year (i.e., $K=350$). For smaller $D$ values, the unlocked capacity of DF  may be lower than that of the CF, as the limited delay window may not be sufficient to accommodate the necessary load increased resulting from shifting power from time $t_k$.  

With $K/T=1\%$, if enforcing a uniform $D$ across all the $K$ intervention events, indeed we may need a relatively large $D$  to reach the CF hosting capacity gain with DF. However,  when we evaluate the dynamic minimal delay requirements for different $t_k$'s depicted in the right panel of Fig.~\ref{fig:delay_info:DM}, we can see  the majority of interventions only require short delays, despite one of the intervention requires 210 minute of delay. In fact, all but 11 out of the 350 interventions require no more than 90 minutes of delay, supporting {\bf KO4}.

\subsection{Theory: Connecting DF and CF}\vspace{-2mm}
We proceed to establish a formal connection between the hosting capacity unlocked by CF and DF. In addition to justifying {\bf KO3}, this also allows us to apply results developed in Section~\ref{sec:theory:CM} to the analysis of DF. 
In this subsection, we use $\cs_{K,D}$ to denote the hosting capacity unlocked by DF with parameter $(K,D)$, and $\cs_K$ for that by CF with parameter $K$. We have the following result. 


\begin{theorem}[Connecting DF and CF]\label{thm:3}
For any $D$ and $K$, $\cs_{K,D} \le \cs_K$. Furthermore, under the assumptions of Proposition~\ref{prop:HCAsoln:DM}, there exists a $D \le T$ such that $\cs_{K,D} = \cs_K$ if and only if for all $k=1, \dots, K$,   \vspace{-.1in}
  \begin{equation}
        \sum_{\tau = t_k}^{T} \left[\cres(\tau) - \cs_K \lbnew(\tau) \right] \ge 0. 
        \label{eq:condition_delay}
         \vspace{-.1in}
    \end{equation}
\end{theorem}

Under the same intervention budget $K$, Theorem~\ref{thm:3} suggests that the hosting capacity gained from \blue{DF} is no more than that from CF. Indeed, with curtailment, we can in theory fully cut the new load under the flexible connection program; however, with the delay model, the amount of load that we can reduce in a time slot depends on how much additional load the next $D$ slots can accommodate. Theorem~\ref{thm:3} also provides a necessary and sufficient condition under which $\cs_{K,D} = \cs_K$ for some $D$. 
To understand this condition, note each term in \eqref{eq:condition_delay} represents the residual capacity for time slot $\tau$ if positive, or the deficit in capacity if negative. For each intervention time $t_k$, the DF can support the CF hosting capacity $\cs_K$ with a large enough $D$, provided that  the largest potential delay window starting from $t_k$, i.e., $\{t_k+1, \dots, T\}$, can accommodate the load increasing due to shifting part of the load from $t_k$. This precisely corresponds to the net surplus in capacity, represented by the sum in \eqref{eq:condition_delay}, being non-negative. 

Condition~\eqref{eq:condition_delay} likely holds in practice, and thus Theorem~\ref{thm:3} justifies {\bf KO3}. This is because of properties of order statistics of dynamic hosting capacity $\cnew(t)$ and the expression of  $\cs_K$ established in Proposition~\ref{prop:HCAsoln:CM}. In fact, the summand in~\eqref{eq:condition_delay} is only negative if $\tau = t_{k'}$ for any $k'=1, \dots, K$. Since we are focusing on the parameter regime where $K\ll T$ (e.g., $K/T=1\% $), most terms in the summation are positive. Furthermore, since $t_k$ are time slots corresponding to lowest values of $\cnew(t)$ or   extreme values of $\lag(t)$ (assuming $\lbnew(t) \equiv 1$), and typical loads have low mean-to-peak ratios, the positive terms in~\eqref{eq:condition_delay} are often  large. Thus, the sum is likely positive. This discussion also offers an intuitive explanation for {\bf KO4}: It is often the case that a few slots following an intervention time $t_k$ can already accommodate the delayed load, and thus the required $D_k$ is likely small for most intervention times.

\section{General Radial Networks}\vspace{-1mm}
\label{sec:network}
This section aims to generalize our copperplate network results to general radial networks. 
We will first show that our methods and theoretical results can be directly extended to the general radial network case by \emph{re-defining dynamic residual and hosting capacities} to account for voltage constraints and network parameters. 
We then conduct a case study by applying our methods to the IEEE 123-bus test feeder. 

\subsection{Theory: Bridging General and Copperplate Cases}\vspace{-1mm}
For solving~\eqref{opt:hca}, the only difference between the copperplate and general network case, is the network constraints embedded in~\eqref{opt:hca:c2}. For the copperplate case, it is reduced to~\eqref{eq:copperplate:cons}. We will show the same holds for the network case, provided that we modify the dynamic hosting capacity defined in Definition~\ref{def:dhc} as follows. 

\begin{definition}[Dynamic Hosting Capacity, Network Case]\label{def:dhc:net}
For each $t\in \mathcal T$, we refer to
\vspace{-2mm}
\begin{subequations}
	\begin{align}
		\cres(t)&:= \min\left\{\overline p_0 \!-\! \lag(t),\! \min_{i=1, \dots, n} \!\frac{v_0\!-\underline v_i \!- 2 \blue{\mathbf Z_i(t)} \bm \ell(t)}{2Z_{i,\inew} }\right\},\label{eq:drc:net}\\
		\vspace{-1pt}
		\cnew(t)&:= \cres(t)/\lbnew(t),\label{eq:dhc:net}
	\end{align}
\end{subequations}
	as the dynamic residual capacity and dynamic hosting capacity for time $t$ at bus $\inew$, respectively, where $\blue{\mathbf{Z}(t)} := \mathbf{R} + \mathbf{X} \, \mathrm{diag}(\blue{\bm \eta(t)})$ and $\blue{\mathbf Z_i(t)}$ denotes the $i$-th row of $\blue{\mathbf{Z}(t)}$. 
	
\end{definition}

With this modified definition, we also define the order statistics $\cnew[s]$ and $t_s$, $s\in \mathcal T$, as in Defintion~\ref{def:dhc} with the updated definition of $\cnew(t)$. 
Equipped with Definition~\ref{def:dhc:net}, the network constraint~\eqref{opt:hca:c2} can be converted into a scalar constraint on the modified new load $\cnew \lbnew(t) + u(t)$ for each $t$, which can be shown to be equivalent to~\eqref{eq:copperplate:cons} as in the copperplate case with our new definition of $\cnew(t)$. As a consequence, results for the general network case mimic that in the copperplate case: 

\begin{theorem}[FA-HCA, General Network]\label{thm:4}
Suppose $\blue{Z_{ij}(t)} \!>\!0$ for all $i,\! j\!=\!1, \!\dots\!,n$.
Then the flexibility-aware hosting capacity at bus $\inew$ in a radial network satisfies Proposition~\ref{prop:HCAsoln:CM}, Proposition~\ref{prop:HCAsoln:DM}, and Theorem~\ref{thm:3}, provided that $\cres(t)$ and $\cnew(t)$ are re-defined as in Definition~\ref{def:dhc:net}.

\end{theorem}

Theorem~\ref{thm:4} states that the methods for solving~\eqref{opt:hca} with both CF and DF can be directly applied to the network case, and the relation between the capacity unlocked by CF and DF does not change. The impact of the network, including its topology, parameters, and voltage constraints, to our analysis is fully encapsulated by the new notion of the dynamic residual capacity~\eqref{eq:drc:net} and dynamic hosting capacity~\eqref{eq:dhc:net}. 

In theory, we can also extend our results in Section~\ref{sec:theory:CM} to the general network case. This would require us to translate our assumptions on $\lag(t)$ to $\cres(t)$ in the copperplate case, and extend such assumptions  to the network case. We leave such exploration to future work. 

\vspace{-2mm}
\subsection{Case Study: IEEE 123-Bus Test Feeder}\vspace{-2mm}
We base our experiments on a single-phase model of a modified IEEE 123-bus radial test feeder  \cite{bobo2020second}, \blue{with $\boldsymbol{\eta}(t)$ following the load power factor at each bus and treated as time-stationary without loss of generality}. Static loads at 85 load buses are used to scale 85 15-minute load time series randomly selected from the LA county in \cite{osti_1854582} to form a test case for a year. To ensure feasibility with respect to the voltage constraints, we then perform a two-step uniform scaling of all loads: a) scaling down the loads so the minimum voltage across the network is equal to $\underline v_i = 0.95$ p.u., with the substation voltage being $1.0$ p.u. and the transformer capacity $\overline p_0$ set as the total real power consumption, and b) further scaling down the loads by 10\% so the network has non-zero residual capacity to support new loads. 
We adopt the same new load profile as in Section~\ref{sec:EO:CM}, located at bus $\inew =13$.

\begin{figure}[h]
\vspace{-.05in}
    \centering
    \begin{subfigure}[b]{\linewidth}
    \centering
	 \includegraphics[width=\linewidth]{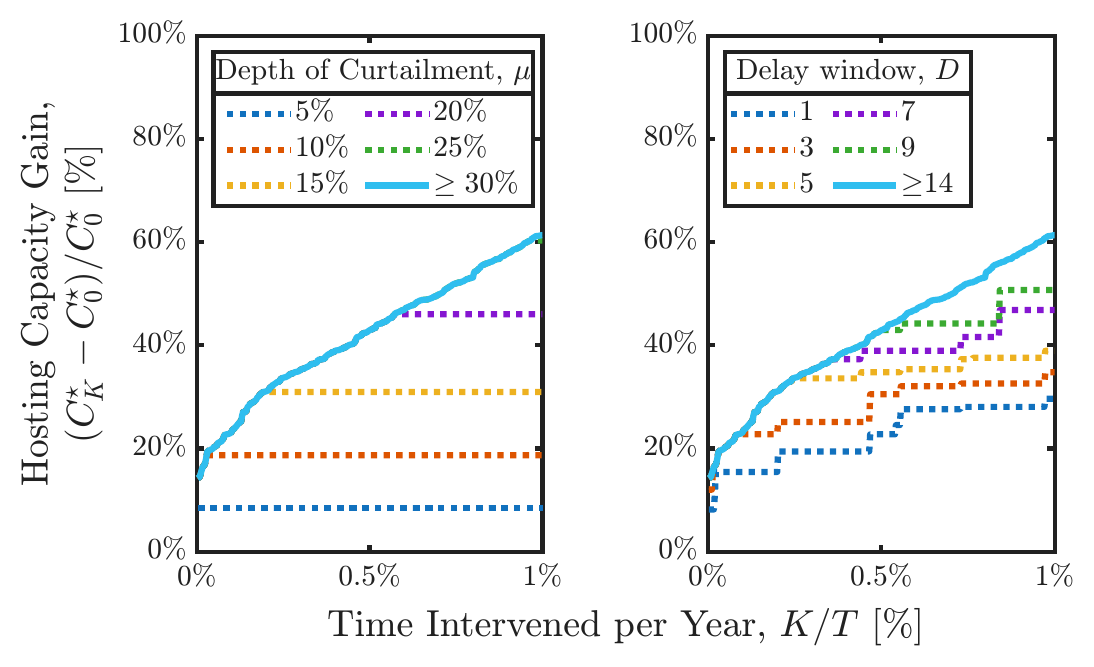}
	 \caption{Hosting capacity gain for CF (left) and DF (right).}
	 \end{subfigure}
	    \centering
    \begin{subfigure}[b]{\linewidth}
    \hspace{0.06in}
	 \includegraphics[width=\linewidth]{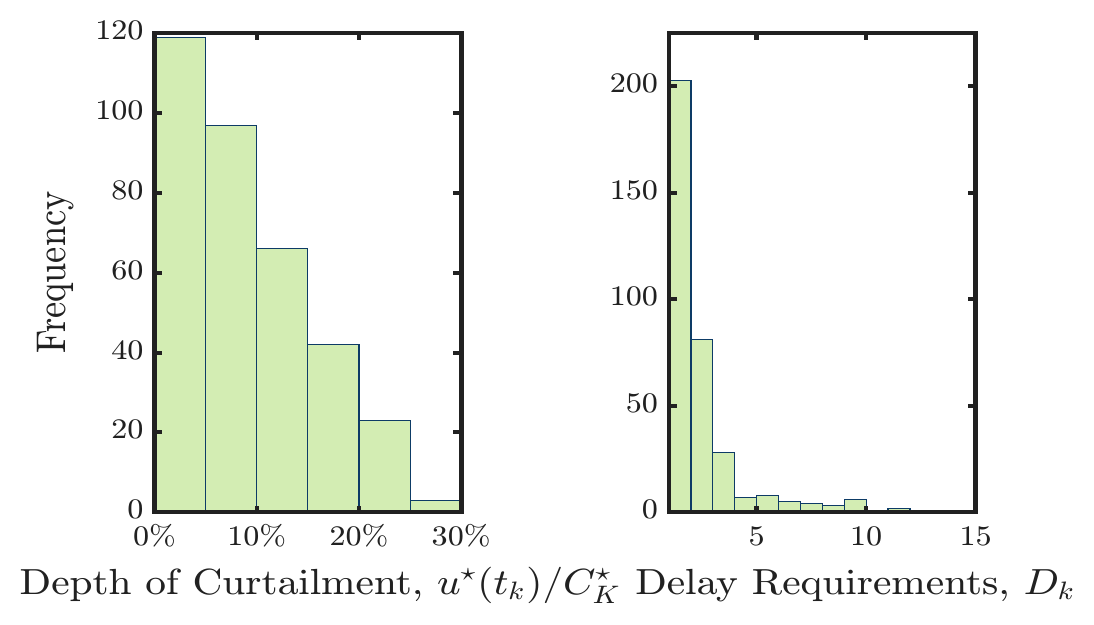}
	 \caption{Depth of curtailment requirements for CF (left) and delay requirements for DF (right) for $K/T=1\%$.}
	 \end{subfigure}
	   \caption{Results for the IEEE 123-bus test feeder.}
    \label{fig:flexible_connection_summary}
    \vspace{-.1in}
\end{figure}

Results are depicted in Figure~\ref{fig:flexible_connection_summary}. Comparing to Figures~\ref{fig:curtail_info:CM} and~\ref{fig:delay_info:DM},
we note that {the} key observations, i.e., {\bf KO1}-{\bf KO4}, still hold for the network case considered. 
\section{Concluding Remarks}\vspace{-2mm}
\label{sec:conclusion}
\vspace{-1mm}
This paper investigates how flexible connection can enhance hosting capacity in distribution networks by allowing controllable loads to be interconnected under infrequent, utility-managed interventions. 
We formulate the FA-HCA problem that explicitly limits the number of allowed interventions, and develop efficient solution methods for both curtailment- and delay-based flexibility models. 
Through empirical testing and theoretical analysis, we show that even a limited number of curtailment or delay events can unlock a significant amount of hosting capacity, requiring {a} small or modest depth of curtailment or delay time for the majority of the infrequent interventions. 
{Our model focuses on planning with historical load profiles, while its usefulness depends on sufficient operational observability to trigger interventions safely. Future work will incorporate uncertainty under partial observability.}

\renewcommand*{\bibfont}{\small}
\setlength\bibitemsep{0.03pt}       
\setlength\bibnamesep{0pt}       
\printbibliography
\clearpage
\end{document}